\documentstyle[epsf,multicol,aps,prl,epsfig] {revtex} 

\newcommand{\be}{\begin{equation}}                                    
\newcommand{\ee}{\end{equation}}                                      
\newcommand{\bea}{\begin{eqnarray}}                                   
\newcommand{\eea}{\end{eqnarray}}                                     
\newcommand{\bsa}{\begin{subeqnarray}}                                
\newcommand{\esa}{\end{subeqnarray}}

\begin{document}                                                      
                                                                      
\title{ Metastability of life }
                                                                     
\author{ V.I. Marchenko }
 
\address{P.L. Kapitza Institute for Physical Problems, RAS,      
117334, Kosygin str. 2, Moscow, Russia}                    
\address{Permanent address. e-mail: mar@kapitza.ras.ru}
\address{Institut f\"ur Festk\"orperforschung, Forschungszentrum      
J\"ulich,                                                           
D-52425 J\"ulich, Germany} 
                                                                      
\vskip 0.2 cm                                                        
\author{\parbox{397pt}{\vglue 0.3cm \small                            
The physical idea of the natural origin of 
diseases and deaths has been presented.
The fundamental microscopical reason is the 
destruction of any metastable state by thermal activation of a nucleus of a
nonreversable change. On the basis of this idea the quantitative theory
of age dependence of death probability has been constructed. The obtained
simple Death Laws are very accurately fulfilled almost for all known
diseases.
}}
\author{\parbox{397pt}{\vglue 0.3cm \small                            
PACS numbers:87.90.+y, 89.90.+n}}
\maketitle                                                            
\begin{multicols}{2}                                                  

All of us will die, as well as all other living organisms and plants.
Each and every machine or construction will breaks. Mountains will fall down or 
earthquakes will happen.

Why? Physics gives the general answer - all of these systems are not in
a full equilibrium. All the systems are metastable, it means: 1) they are 
stable against small external influences, but 2) each of them, as the
worst ones, as well as the best ones, has a finite probability to be 
spontaneously destroyed without any external influence even in the ideal 
environment and at the perfect conditions. According to Gibbs \cite{LL-S} the 
fundamental reason of the destruction is the thermal activation of 
critical nucleus of nonreversable change in the system.

Let us consider a simple example - a stretched ideal monocrystal string.
If we wait sufficiently long time the temperature fluctuations will 
produce a critical  Griffith's crack \cite{Grif} at some place and the 
string will break. It is possible that the critical crack will appear 
earlier if there are some defects in the crystal. Such a nucleation 
process occurs in different ways for different cases (activation of point 
defects in the crystals, condensation in a super saturated solution, 
nucleation of a new phase in a first order phase transition) and it is 
well studied in condensed matter physics.

Any living organism is a much more complicated system, but the described
phenomena should occur in it also. The thermal activation of critical 
nucleus is the last and unremovable killer. Last - if we exclude all 
other origins of diseases and deaths. Unremovable, but, one can hope - 
not untreatable.
 
I want to stress here that the known qualitative and quantitative facts
about majority of diseases can be understood from the point of view of 
theoretical physics in terms of metastability and activation of critical 
nucleus. So, I do think that the thermodynamic killer works, and it is 
the main killer.
 
Gompertz \cite{G} discovered that a probability {$D(x)$} to die at 
the age {$x$} in the time interval $dt$ exponentially increases with 
age 
\be
D\propto exp\left(\frac{x}{a}\right).
\ee
According to modern mortal statistics Gompertz law is valid at the age 
range $30\div 70$ years, and even more strong increase appears at older 
ages. Exponential age dependence of $D$, from my point of view, is the most 
crucial sign on the nature of micro origin of diseases leading to death. 

I have no answer for many questions one can ask about details of the 
relationship between a given disease and the proposed idea of their natural 
micro origin. Only I can do for the moment is to present a theory of 
age dependence of probability of arising of the nucleus. 
 
On a molecular (and macromolecular) level there are few reasons of arising 
of almost non removable point defects, for example, due to the process of 
oxidation \cite{N}. Thermal fluctuations should produce configurational 
transformations of individual molecules \cite{SH}. The same effect can be 
caused also by some external agents (photons, impurity atoms or molecules, 
elementary particles). If a concentration of those point defects is small,
then the probability of arising of new defects does not depend upon the 
interaction between them. It means that the concentration of point defects 
should be simply proportional to the age $x$. This linear law is known in
an absolutely analogies situation, Zeldovich stage of nucleation in I order 
phase transition \cite{LL-K}. It is quite natural to 
assume, that at any age the dimensionless molecular concentration of the 
point defects remains small, so at any age this law is valid. 
 
Growing concentration of the point defects gives rise to small changes of 
physical parameters of body structures on a macroscopic scale (membranes, 
cells, as well as on a higher levels). One can imagine that some 
functionally significant defects are thermally activated on this scale 
(example, arising of Griffith-like critical crack in a micro cappilary, 
periodically stressed by oscillating blood pressure) or point defects 
tend to precipitate into a condensed state (as it is in supersaturated 
solutions), or even some type of a structural phase transition occurs at 
some critical value of the defect concentration. Some of such types of 
spontaneous changing in the body can have serious functional consequences 
leading to diseases, and death.

The probability $W$ of arising of such micro damages is governed by Gibbs 
law 
\be
 W\propto exp\left(-\frac{U}{T}\right),
\ee
where $U$ is the minimum energetic barrier of the unreversible change 
(critical nucleus), and $T$ is the temperature. Usually it is possible to 
expand energy of critical nucleus in the small concentration, or 
equivalently in age: $U=U_0 + U'x$, and if $U'$ is negative, the 
barrier diminishes with the age, we obtain the exponential law, Eq.(1).
If $U'$ is positive, one has the growth of the barrier, and the stability 
of the body increases. It is possible that the age decreasing of the
infant mortality is partly related to this circumstance.
 
The expansion of $U$ in concentration is impossible in the case of
condensation in a supersaturated gas with small concentration (as well as 
in the vicinity of I order phase transition). In a two-dimensional 
condensation of supersaturated gas the energy of the critical nucleus is 
inversely proportional to the concentration, or in our case 
$U \sim x^{-\small 1}$,  corresponding to the second exponential 
law
\be
W\propto exp\left(-\frac{b}{x}\right).
\ee
In a three-dimensional condensation there should be 
$U\sim x^{-\small2}$, and the third exponential law is
\be
W\propto exp\left(-\frac{c}{x^2}\right).
\ee
 
Let us consider the US-97 death statistics specified by selected 
causes \cite{FD}. If one plots $ln(D_i)$ v.s. $x$, or, 
v.s. $1/x$, and $1/x^2$ it is easy to find that almost all cases have a clearly 
distinguishable age behaviors: 20 cases of Gompertz exponential law, Eq.(1); 
14 cases of second exponential law (3); 4 cases with more complicated 
behavior, but the laws (1) or (3) are valid there in a wide age range, and 
some strange crossover occurs to some other behavior; 24 cases are 
not related with aging. Only in 3 cases statistics does not permit 
to make a definite conclusion on the type of the age dependence. 
Examples of the clearly detactable exponential age behavior of death rate 
presented in Fig.1-4. 

Death rate here is the number of 1997 year deaths per 100.000 
population of specified age groups 0-5, 5-14, ... 75-84, 85 years and over.
There are a lot of intriguing coincidences of parameters  $(a,b)$ 
for different diseases. It possibly means, that a number of discussed different 
micro origins is substantially smaller than a number of diseases. Some of 
diseases arise  presumably as a combined effect of two 
different micro origins. This analysis is in progress.

\begin{figure} 
\begin{center}
\epsfig{figure=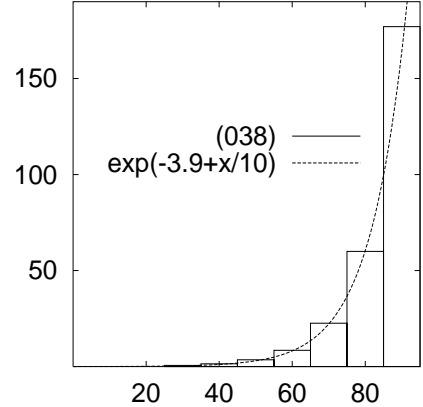, width=6cm,height=6cm,angle=-90}
\caption{Septicemia (038). Death rate }
\end{center}
\end{figure} 

\begin{figure} 
\begin{center}
\epsfig{figure=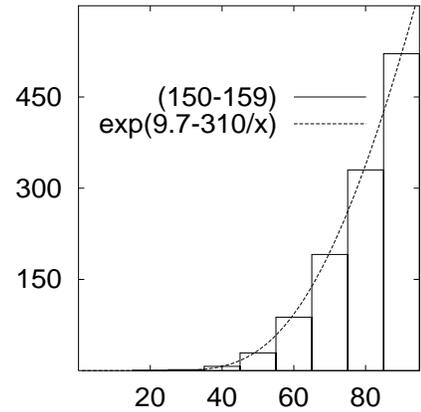, width=6cm,height=6cm,angle=-90}
\caption{Malignant neoplasms of digestive organs and peritonium (150-159). Death rate.}
\end{center}
\end{figure} 

\begin{figure} 
\begin{center}
\epsfig{figure=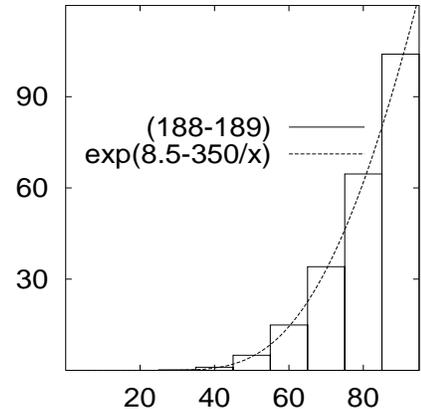, width=6cm,height=6cm,angle=-90}
\caption{Malignant neoplasms of urinary organs (188-189). Death rate.}
\end{center}
\end{figure}

\begin{figure} 
\begin{center}
\epsfig{figure=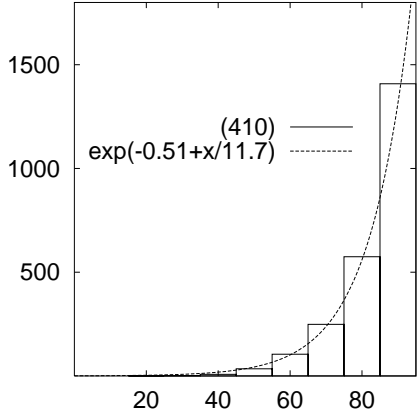, width=6cm,height=6cm,angle=-90}
\caption{Acute myocardial infarction (410). Death rate.}
\end{center}
\end{figure}

The characteristic magnitude of function D in cases with Gompertz law (1) 
at $x=0$ is $exp(-13\div-22)$ per year, or $exp(-30\div-39)$ per second. 
Let us compare this value with Eq.(2). One should introduce some pre-exponent 
value. Its most simple estimate is the characteristic frequency of
oscillations of atoms in condensed matter $\omega\sim k\theta/\hbar$, 
where $\theta\sim 10^2 K$ is a Debye temperature, $k$ - Boltzmann's 
constant, $\hbar$ - Planck's constant. One should introduce an 
additional factor, an effective number N of possible places where 
the given critical nucleus can arise. The temperature of the body 
is $T=273+36.6\approx 310 K$. The comparison gives a reasonable 
estimation of barriers $U\sim(1.2\div1.4)*10^4 K+TlnN$, or
$U\sim1.1\div1.3 eV$ if $N\sim 1$, and only $U\sim 3 eV$ even if N 
is equals to total amount of molecules in a body, this effective number 
is of course unrealistic, and I want just to note here that in any case 
the barrier estimation gives value usual in condensed matter physics.

In order to estimate the age change of barriers one does 
need not to know the pre-exponent factor in the expression (2).
Typical 90 years increasing factor of $D_i$ is $exp(8)$. It corresponds 
to diminishing  of barriers $\delta U \sim 8T$, this value is also 
reasonable $\delta U\sim0.2 eV\ll U$. Two parameters, the small one  
$\delta U/U\ll 1$, and the big one $U/T\gg 1$, are the main parameters
of the theory.  

In the framework of presented picture the small difference in barriers  
of the order of $0.02eV$ for male and female corresponds to known ratio 
$D_m/D_f\sim 2$, and  can be directly related to the difference 
1/23 in chromosome compositions. The variation 
of parameters on time, and specific groups of population, countries, races, etc., 
should be of the same order of magnitude. The situation is 
similar to the usual one in condensed matter physics,  where experimental 
data are observably dependent on sample preparation conditions.

Note, that there is no real contradiction between presented idea and the 
fact that there is a lot of diseases cased by viruses and bacteria. The 
age dependence of those diseases should be related to some micro origin 
of the destruction of the immune system.  
 
I think also, that discussed thermal activations should play not the last 
role in a generation of congenital anomalies.

                        

\end{multicols}                                                       
\end{document}